# Title Page

Author:
Dr.-Ing. Barış İRHAN

Title:
A universal predictor-corrector type incremental algorithm for the construction of weighted straight skeletons based on the notion of **deforming** polygon

Company:
Arkitech Advanced Construction Technologies

Address:
Malıköy Başkent OSB Mahallesi Atatürk Bulvarı No: 23
Sincan / Ankara / Turkey

E-mail:
barisirhan@arkitech.com.tr

Tel:
+90 534 210 81 70

# Abstract

A new predictor-corrector type incremental algorithm is proposed for the exact construction of weighted straight skeletons of 2D general planar polygons of arbitrary complexity based on the notion of **deforming** polygon. Construction of weighted straight skeletons for planar polygons is just a particular realization of the underlying algorithm. Algorithm is general, sound, simple and easy to implement. It takes its roots from the techniques which are commonly used in computational solid mechanics.

In the proposed algorithm, the raw input provided by the polygon itself is enough to resolve edge collapse and edge split events. Neither the construction of a kinetic triangulation, for the regions of the polygon which are not yet swept, nor the computation of a motorcycle graph, or any other means, is required. Polygons with weighted, positively or negatively, or stationary edges are treated within the same framework. No differentiation is made between convex and reflex vertices because they lose their dictionary meanings for a polygon exposed to an arbitrary deformation process. Due to its incremental nature, there is always a room in the algorithm for the interactive construction of the straight skeleton. For example, the edges which start to move at different times could easily be handled or the amount of offset applied could be manually adjusted. Last but not the least, degenerate events, where simultaneous edge collapse or edge split events take place at the same location at the same time, are tackled in a robust elegant way.

The proposed algorithm is of predictor-corrector type. In the algorithm, the edge collapse and edge split events are tackled by a completely different novel original approach which is first of its kind. In the predictor step, the position of the vertices is advanced in time by direct integration assuming no event. Then predicted positions are corrected by using linear interpolation if there are edge collapse or edge split events within the same increment. In the algorithm edge collapse and edge split events are detected by, respectively, edge swap and edge penetration. Both edge swap and edge penetration have counterparts in computational solid mechanics. Former statement is motivated by impenetrability of matter which pose a constraint on the deformation gradient such that its determinant, Jacobian, must be sign preserving. For one dimensional case it means that projected length of a segment must preserve its sign. Latter statement is motivated by kinematical constraint posed when two deformable body, or body with itself, come into contact. It simply states that bodies cannot penetrate into each other during contact.

The proposed algorithm has been used to construct roof topology starting from a floor plan of various complexity ranging from simple convex to highly nonconvex with holes. In order to construct, improve and test the building blocks of the underlying algorithm, a graphical user interface, Straight Skeleton Development Kit, has also been developed in parallel by the author using C++ programming language.

# 1 Introduction

Computational solid mechanics is a subdiscipline of computational mechanics where the partial differential equations governing the evolution of physical quantities, like linear momentum, heat (energy), mass, etc., are integrated, in space and time, by resorting to vast amount of numerical techniques available [1–3].

For subsequent developments, we will investigate set of evolution equations for linear momentum which is governed by a second order hyperbolic partial differential equation and has the form

$$\rho \dot{\boldsymbol{v}} = \boldsymbol{\nabla} \cdot \boldsymbol{\sigma} + \rho \boldsymbol{b} \tag{1}$$

where $\rho$, $\boldsymbol{v}$, $\boldsymbol{\nabla}$, $\boldsymbol{\sigma}$ and $\boldsymbol{b}$ are, respectively, mass density, velocity, gradient vector, stress tensor and body force associated with the material point. Eq.(1) simply states that temporal evolution of linear momentum at a material point is equal to sum of the net momentum flux (force acting at the vicinity like contact) and the source term (force acting at a distance like gravity) associated with the material point. This equation is called as balance equation for linear momentum, momentum equation or wave equation in the literature and in general terms it governs the propagation of stress waves in inelastic media. Note that momentum equation is nothing else but the continuum generalization of Newton's second law of motion, $f = ma$.

In computational solid mechanics, Lagrangian description of the motion is generally adapted in order to track the evolution of physical processes. In Lagrangian framework, the motion of each material particle is tracked throughout the course of the deformation process and the existence of a motion function of the form

$$\boldsymbol{x} = \boldsymbol{\phi}(\boldsymbol{X}, t) \tag{2}$$

is postulated. In Eq.(2), $\boldsymbol{x}$ and $\boldsymbol{X}$ are, respectively, the position of a typical material point at initial and current configuration of the body. In order to be able to define the deformation of a solid, the motion function $\boldsymbol{\phi}$ must possess some properties. To analyze the deformation locally around the material point, use of the gradient of the motion function, also called as deformation gradient, is made. It has the following standard form:

$$\boldsymbol{F}(\boldsymbol{X}, t) = \frac{\partial \boldsymbol{\phi}}{\partial \boldsymbol{X}} = \boldsymbol{\nabla} \boldsymbol{\phi} \tag{3}$$

In order to define the solid deformation, the following conditions must hold for $\boldsymbol{F}$ [4]:

$$\boldsymbol{F}d\boldsymbol{X} \neq \boldsymbol{0} \quad \forall \quad d\boldsymbol{X} \neq \boldsymbol{0}$$
$$\boldsymbol{F}d\boldsymbol{X} = \boldsymbol{0} \quad \leftrightarrow \quad d\boldsymbol{X} = \boldsymbol{0} \tag{4}$$

From physical standpoint above conditions state that an infinitesimal material line element with non-zero length cannot be reduced down to a zero-length line element during deformation. This constraint implies that $\boldsymbol{F}$ must be non-singular and therefore has non-zero determinant, i.e. $\det\boldsymbol{F} \neq 0$. The volume of a typical material point at current configuration is given by the expression

$$dv = \det\boldsymbol{F}\, dV \tag{5}$$

where $dV$ is its volume at initial configuration. For a solid, the volume of the material point must be positive throughout the course of the deformation process. Therefore, in addition to being nonzero, the determinant of the deformation gradient must also preserve its sign, i.e.,

$$\det\boldsymbol{F} > 0 \tag{6}$$

The inequality constraint given in Eq.(6) is also known as the impenetrability of matter. Inability of the motion function given by Eq.(2) to satisfy this constraint is a clear indication of severe deformations. The parts of the body experiencing severe deformations does not behave like a solid anymore but a fluid or a gas. Standard techniques fall short when a solid starts to flow or a fluid starts to deform. The possible change of the material state, which could also be called as change in topology, is the driving force behind several discretization techniques, like ALE [5, 6], adaptive remeshing [7–10], adaptive element deletion [11], coupled SPH-FEM [12–14], etc. if one insists to stay in the Lagrangian framework throughout the course of the deformation process.

Predictor-corrector type numerical algorithms are commonly used to solve problems where discontinuity, change from one state to another, is expected in the solution. For example, in multibody contact problem the solution of the momentum equation, Eq.(1), must also respect the kinetical and kinematical contact constraints [15] which are defined as

$$g_n \geq 0; \quad t_n \leq 0; \quad t_n g_n = 0 \tag{7}$$

where, $g_n$, $t_n$ are, respectively, normal component of the gap vector and normal component of the contact traction. These inequality constraints imply that during contact bodies cannot penetrate into each other, $g_n \geq 0$, and normal component of the contact traction transmitted must be compressive, $t_n \leq 0$. As mentioned before, multibody contact problem described could be solved by employing a predictor-corrector type algorithm. Employing, for example, finite element method [1–3] to discretize momentum equation in space, the discrete dynamic equilibrium equation for a typical node *I* is written as

$$m_I \boldsymbol{a}_I = \underbrace{\boldsymbol{f}_I^{ext} - (\boldsymbol{f}_I^{str} + \boldsymbol{f}_I^{con})}_{\boldsymbol{f}_I} \tag{8}$$

where $m_I$, $\boldsymbol{a}_I$, $\boldsymbol{f}_I^{ext}$, $\boldsymbol{f}_I^{str}$ and $\boldsymbol{f}_I^{con}$ are, respectively, lumped mass, acceleration, external force, internal force and contact force associated with typical node *I*. For high velocity short duration impact problems, Eq.(8) is usually integrated in time by resorting to an explicit time integration scheme [7, 8, 11]. Using, for example, second order central differences, nodal acceleration at time step $t_n$ can be approximated as

$$\boldsymbol{a}_{In} \cong \frac{\Delta \boldsymbol{x}_{In} - \Delta \boldsymbol{x}_{In-1}}{\Delta t^2} \tag{9}$$

For brevity, we drop subscript *I* appended for the derivations that follow. Combining Eq.(8) and Eq.(9), an explicit expression for the position update is then obtained as

$$\Delta \boldsymbol{x}_n = \frac{\Delta t^2}{m} \boldsymbol{f}_n + \underbrace{\Delta \boldsymbol{x}_{n-1}}_{\text{known}} \tag{10}$$

In the presence of contact interactions, unfortunately, position update cannot be performed in single step because contact forces are not known beforehand. One can remedy such a situation by setting up a predictor-corrector type algorithm. In the algorithm, $\Delta \boldsymbol{x}_n$ is first additively decomposed as

$$\Delta \boldsymbol{x}_n = \Delta \boldsymbol{x}_n^p + \Delta \boldsymbol{x}_n^c \tag{11}$$

where $\Delta \boldsymbol{x}_n^p$, $\Delta \boldsymbol{x}_n^c$ are, respectively, predictor (known) and corrector (unknown) position increments. In the predictor step node-level update is performed assuming no contact. Therefore, for typical node *I* we have

$$\boldsymbol{f}_n^p = \boldsymbol{f}_n^{ext} - \boldsymbol{f}_n^{int} \tag{12}$$

$$\boldsymbol{f}_n^{con} \overset{!}{=} 0$$

where $\boldsymbol{f}_n^p$ is the total predictor (known) force acting on node *I*. Using Eq.(10), predictor position increment and predictor position itself are obtained as

$$\Delta \boldsymbol{x}_n^p = \frac{\Delta t^2}{m} \boldsymbol{f}_n^p + \Delta \boldsymbol{x}_{n-1} \tag{13}$$

$$\boldsymbol{x}_{n+1}^p = \boldsymbol{x}_n + \Delta \boldsymbol{x}_n^p$$

Next, based on predictor position given by Eq.(13), a global contact search is performed, and it is checked if there are some nodes for which impenetrability constraint given by Eq.(7) is violated. If there is no violation, predicted positions are correct and we have

$$\boldsymbol{x}_{n+1} = \boldsymbol{x}_{n+1}^p \tag{14}$$

On the other hand, if there is violation, one has to take another step and compute corrector position increment $\Delta \boldsymbol{x}_n^c$ due to unknown contact force $\boldsymbol{f}_n^{con}$. Assuming that the contact force is known, the corrector position increment $\Delta \boldsymbol{x}_n^c$ is then obtained as

$$\Delta \boldsymbol{x}_n^c = -\frac{\Delta t^2}{m} \boldsymbol{f}_n^{con} \tag{15}$$

Unknown contact force $\boldsymbol{f}_n^{con}$ can be computed by exploiting the kinematical contact constraint given in Eq.(7), see [11] for details. Note that predictor-corrector type algorithms are also commonly used for the numerical integration of inelastic constitutive equations [16–18].

The inequality constraint given by Eq.(6), namely impenetrability of the matter, and the realization of predictor-corrector type algorithm for the solution of multibody contact problems motivate a new original algorithm for the construction of weighted straight skeletons of planar polygons.

To make a transition to the straight skeleton, consider the closed planar polygon given in Fig. 1, where its configuration at different time instances is depicted. In classical straight skeleton problem, each edge of the polygon moves inwards with a constant unit velocity in such a way that it remains essentially parallel to itself throughout the course of the deformation process, which continues until the polygon shrinks down to a segment or a point. The regions swept by the polygon edges throughout the process construct faces of the straight skeleton structure of the underlying polygon. During the deformation process, polygon topology might change due to edge collapse and edge split events. The edge collapse event takes place at an instance when at least one edge of the polygon collapses down to a point, whereas the edge split event comes into play as at least one of the vertices hits an edge or another vertex of the polygon. Velocity of the vertices involved changes as these events happen and change the topology. Therefore, these events must be detected and handled properly in order to end up with correct straight skeleton structure.

First algorithm to construct straight skeleton of polygons was appeared in [19]. It is based on wavefront propagation. In the algorithm, a priority queue with respect to the occurrence time of edge events is constructed and processed. To determine split events, it is supplemented by a triangulation algorithm. In [20, 21], an algorithm similar to one given in [19] was presented with a different strategy to handle reflex vertices. In [22], use of kinetic triangulation is made to detect edge collapse and edge split events. Events are signaled as soon as a triangle collapses down to a segment or a point. In case of triangle flip, however, triangle collapse is not always associated with an edge collapse or edge split event. Therefore, edge flip events must be detected and filtered out which brings additional overburden and shadows the simplicity of the underlying idea. Also note that in the algorithm solution of a quadratic function is required to determine the collapse time of a triangle which brings additional cost and possible loss of accuracy together. In [23, 24], an approach to eliminate flip events based on the Steiner triangulations was proposed and based on this observation a straight skeleton algorithm for arbitrary planar straight line graphs based on generalized motorcycle graph was presented. This work was later extended in [25] to construct straight skeletons of positively weighted polygons based on bisector arrangement.

(a)

(b)

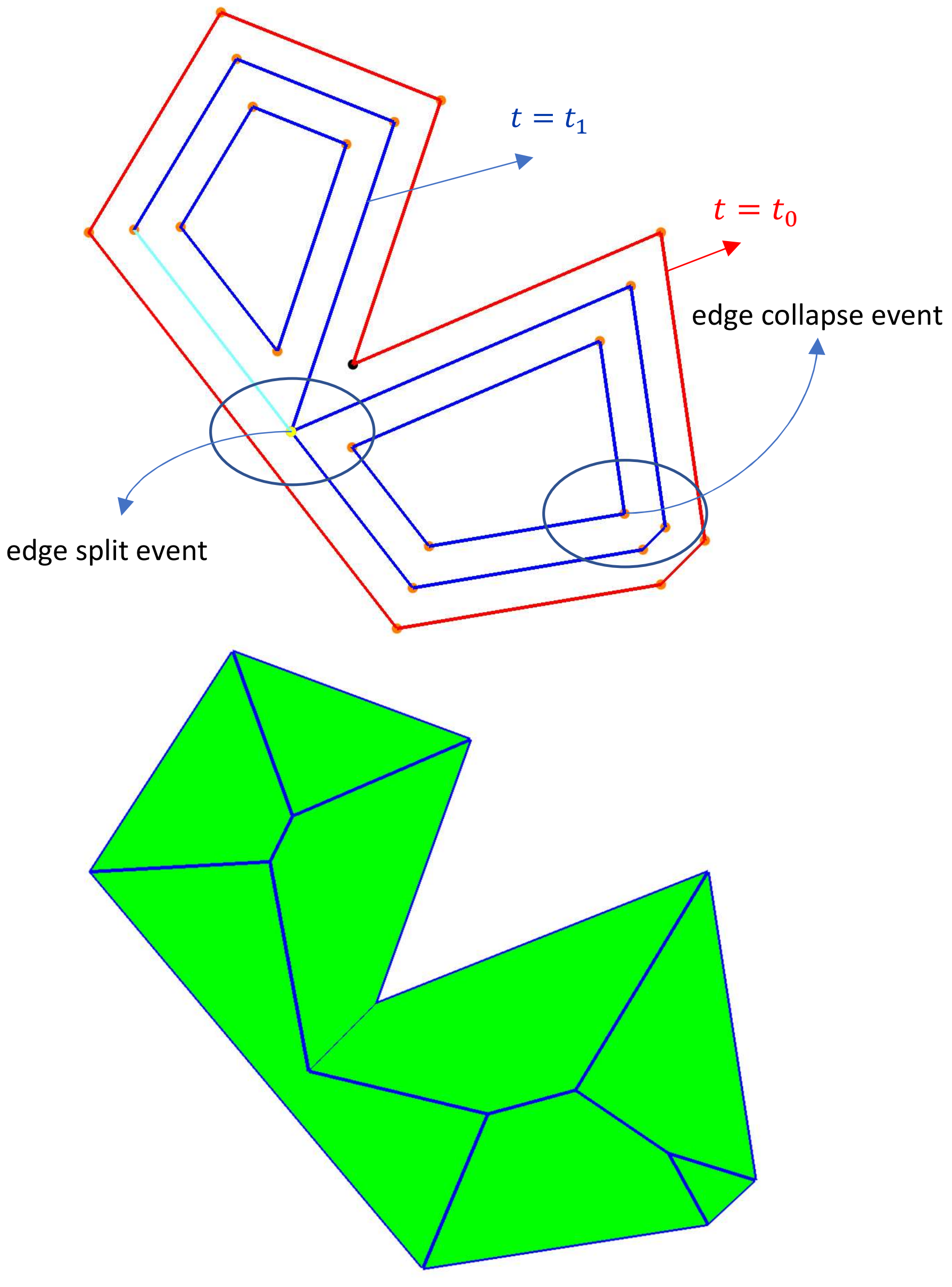

*Fig. 1 (a) Deforming polygon. (b) Straight skeleton structure*

In [26], the classical algorithm was enhanced by enabling the propagation of the edges at different speeds and hence the weighted version of the straight skeleton algorithm was emerged. The classical and positively weighted version of the algorithm are similar. However, there is a price to pay for the additional flexibility offered by weighted version. Due to additional degree of freedom provided, some ambiguities might arise as the edge collapse event happens. After processing such an event, one might end up with vertices which are colinear. For a colinear vertex, the set of equations governing its propagation velocity turns out to be linearly dependent and hence the solution is undefined. Therefore, after processing edge collapse events, colinear vertices must be found and removed from the forefront polygon. However, removing the colinear vertices is not enough. One has to take another step and decide how the region, line segment, associated with a colinear vertex will continue its propagation [27, 28]. Note also that both the classical and the weighted version of the algorithm might suffer from the ambiguities which arise after detecting and processing multiple vertex events [27–29]. Before passing note that it has been shown systematically in [27] that weighted straight skeleton have different geometrical characteristics than its unweighted counterpart.

Motorcycle graph was first introduced in [26] to address the essential problems related with reflex vertices while constructing the straight skeletons. The relationship between motorcycle graph and construction of the straight skeletons was first exploited in [30]. Later on several straight skeleton algorithms [23, 25, 31, 32] based on the motorcycle graph have been published. There are also some other algorithms in the literature for the construction of straight skeletons. In [33], an algorithm for monotone polygons is proposed, whereas in [34, 35], algorithms for simple rectilinear polygons are presented.

Weighted straight skeleton algorithm has a wide range of application area. It has been used, (i) to construct roof and terrains [19, 22, 26, 28, 34, 36–38], (ii) to compute mitered offset curves [39, 40], (iii) in mathematical origami and the fold-and-cut problem [41, 42], (iv) for contour interpolation [43], (v) for polygon decomposition [44], (vi) for digital pathology to improve image segmentation and image analysis [45].

Straight skeleton algorithm has been implemented into several software packages [21, 46, 47]. Of course, without such a tool at hand it is very unlikely to end up with a reliable straight skeleton algorithm.

# 2 Contribution

To the author knowledge, a predictor-corrector type incremental algorithm has never been used before for the construction of straight skeletons. In this work an algorithm of this class is presented. In addition to being original, the algorithm enjoys some very powerful features. First of all, it is simple, intuitive and easy to implement. It is general that is it can be applied to any kind of closed planar polygons of arbitrary complexity equally well. There is no room for exceptional cases. Algorithm works with raw input provided by the underlying polygon. Construction of a kinetic triangulation or a motorcycle graph, etc. is not necessary. Extension to general planar straight-line graphs seems to be straightforward. In this work, speed of the underlying algorithm has not been studied. In its current version it is slow. To construct, improve and test the building blocks of the algorithm, a graphical user interface, Straight Skeleton Development Kit, has been developed in parallel by the author using C++ programming language starting from scratch.

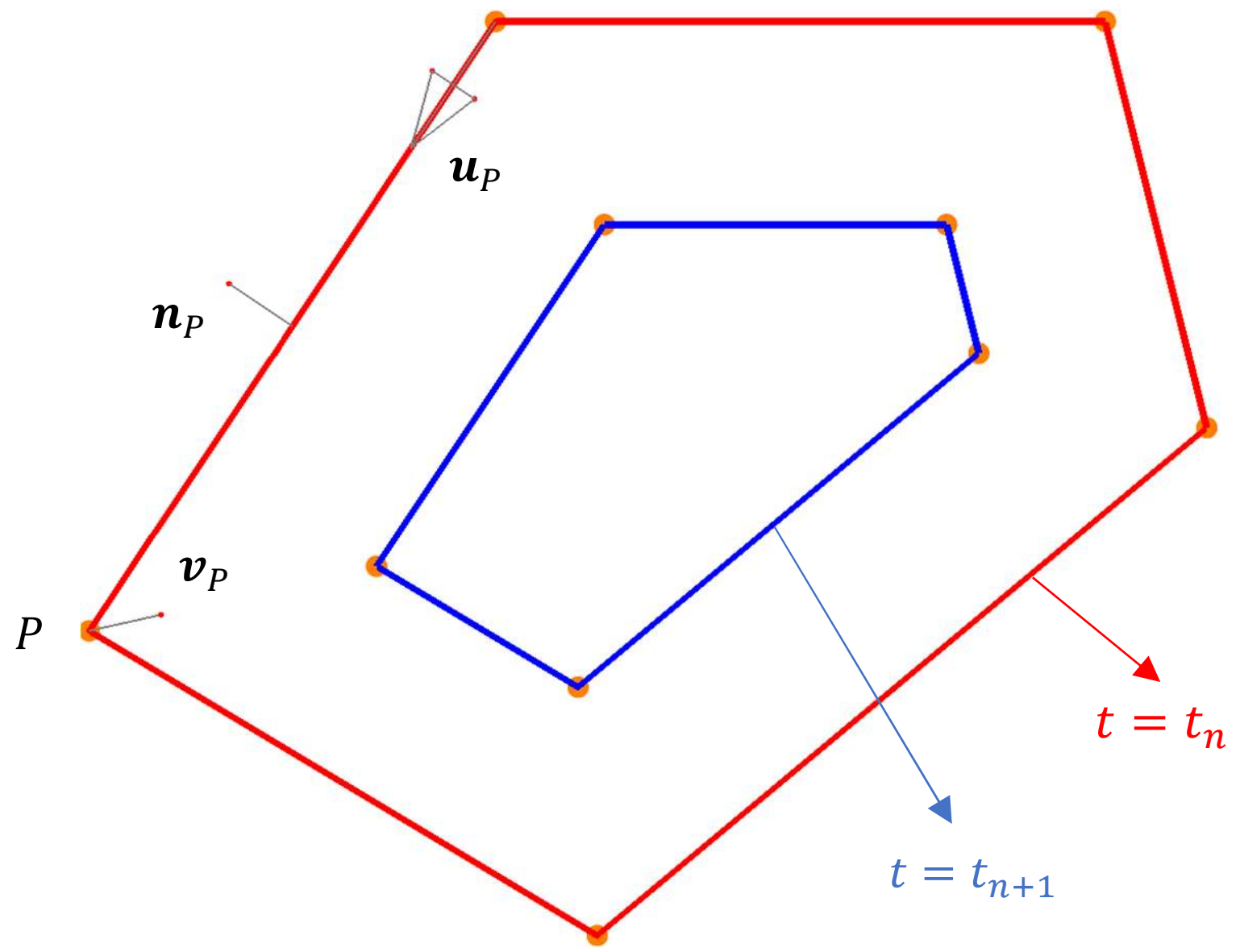

*Fig. 2 A polygon and a typical vertex*

# 3 The algorithm

Consider the closed planar polygon given in Fig. 2, where its configuration at time $t_n$ and $t_{n+1}$ are depicted. Each vertex of the polygon changes position in time with a prescribed constant velocity. Recall that velocity of the vertices has a particular form in case of straight skeleton problem such that an edge belonging to the polygon remains essentially parallel to itself during incremental motion. Then for a given time increment

$$\Delta t_n = t_{n+1} - t_n \quad (16)$$

the position of a vertex at time $t_{n+1}$ can exactly be computed by direct integration as

$$\boldsymbol{P}_{n+1} = \boldsymbol{P}_n + \Delta t_n \boldsymbol{v}_P \quad (17)$$

While the polygon is deforming, the physical constraints given by Eq.(6) and Eq.(7) must be respected. According to impenetrability of matter, see Eq.(6), polygon edges cannot swap during incremental motion. If edge swap happens, it indicates that there are edges which collapse within the given time increment. Second physical constraint is the so-called contact impenetrability, see Eq.(7). That is during deformation, polygon cannot penetrate into itself, from inside to outside or vice versa. If such a penetration occurs, it indicates that there are edges which split within the given time increment. Position of the vertices computed by direct time integration is then used as a predictor if edge collapse or edge split events take place during incremental motion. In such a case, predicted positions must be corrected. Correction phase requires the determination of the exact time instances of the edge collapse and edge split events.

## 3.1 Edge collapse event

Edge collapse event within a time increment is detected by edge swap. An edge is swapped if it changes direction (see Fig. 3).

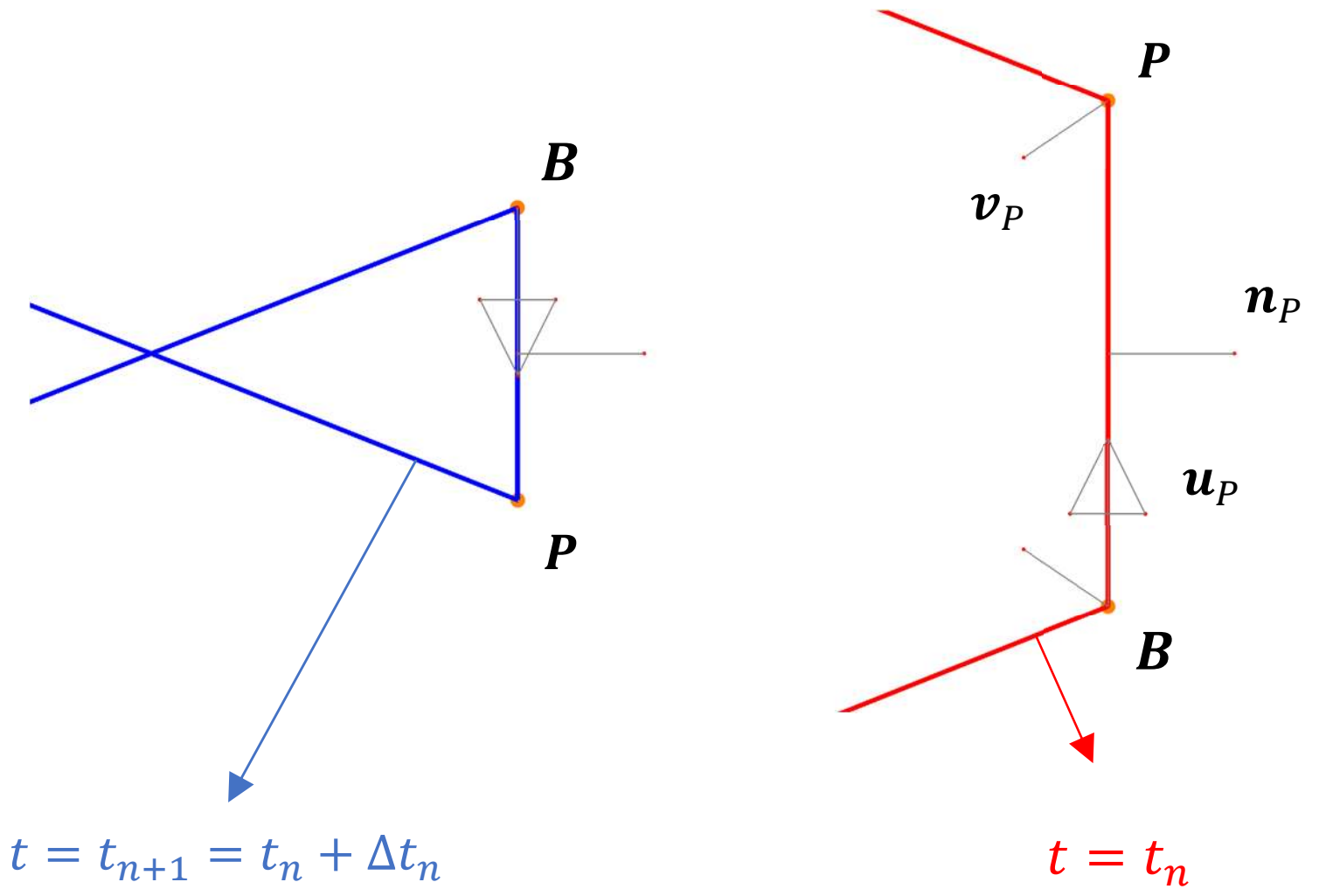

*Fig. 3 Edge swap*

Since vertices move along a linear path with a constant velocity, the instance of an edge collapse event can exactly be computed by linear interpolation. For this purpose, we define relative position vectors for a given vertex *P*, using the vertex *B* before it, corresponding to time steps $t_n$ and $t_{n+1}$ as

$$\boldsymbol{dP}_n = \boldsymbol{P}_n - \boldsymbol{B}_n = l_n \boldsymbol{u}_P \qquad (18)$$
$$\boldsymbol{dP}_{n+1} = \boldsymbol{P}_{n+1} - \boldsymbol{B}_{n+1} = l_{n+1} \boldsymbol{u}_P$$

where projected lengths, $l_n$ and $l_{n+1}$ are given as

$$l_n = \boldsymbol{dP}_n \cdot \boldsymbol{u}_P \qquad (19)$$
$$l_{n+1} = \boldsymbol{dP}_{n+1} \cdot \boldsymbol{u}_P$$

Then one can parameterize projected length within a time increment as

$$l = \frac{1}{2}(1-\xi)l_n + \frac{1}{2}(1+\xi)l_{n+1} \tag{20}$$

with $\xi \in [-1,1]$.

At the instance of edge collapse event the following equation must hold:

$$l = 0 \tag{21}$$

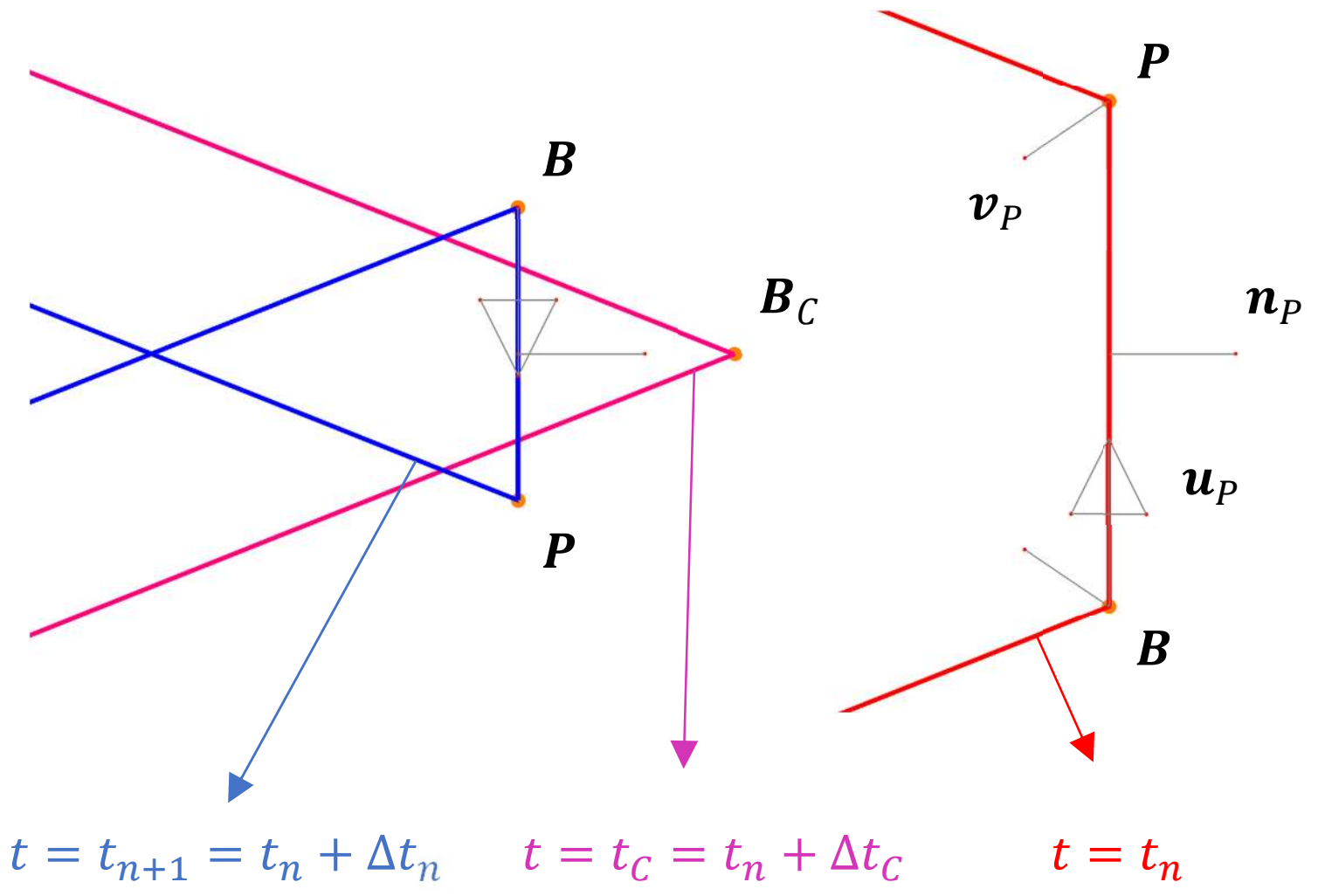

*Fig. 4 The instance of edge collapse event*

Using Eq.(20), local coordinate corresponding to edge collapse can then be computed as

$$\xi_C = -\frac{l_M}{l_D} \tag{22}$$

where $l_M$ and $l_D$ are defined as

$$l_M := \frac{1}{2}(l_{n+1} + l_n) \tag{23}$$
$$l_D := \frac{1}{2}(l_{n+1} - l_n)$$

One can also parameterize the time increment, similar to Eq.(20), as

$$\Delta t = \frac{1}{2}(1-\xi)0 + \frac{1}{2}(1+\xi)\Delta t_n \tag{24}$$

By inserting $\xi_C$ computed in Eq.(22), time increment corresponding edge collapse event, see Fig. 4, is obtained as

$$\Delta t_C = \frac{1}{2}(1+\xi_C)\Delta t_n \tag{25}$$

## 3.2 Edge split event

Edge split event within a time increment is detected by edge penetration (see Fig. 5). Since the vertices move along a linear path with constant velocities, exact instance of edge split event can again be computed by using linear interpolation. For this purpose, we define parameterization of vertices $\boldsymbol{P}$ and $\boldsymbol{T}$ as

$$\boldsymbol{P} = \frac{1}{2}(1-\xi)\boldsymbol{P}_n + \frac{1}{2}(1+\xi)\boldsymbol{P}_{n+1}$$
$$\boldsymbol{T} = \frac{1}{2}(1-\xi)\boldsymbol{T}_n + \frac{1}{2}(1+\xi)\boldsymbol{T}_{n+1} \quad (26)$$

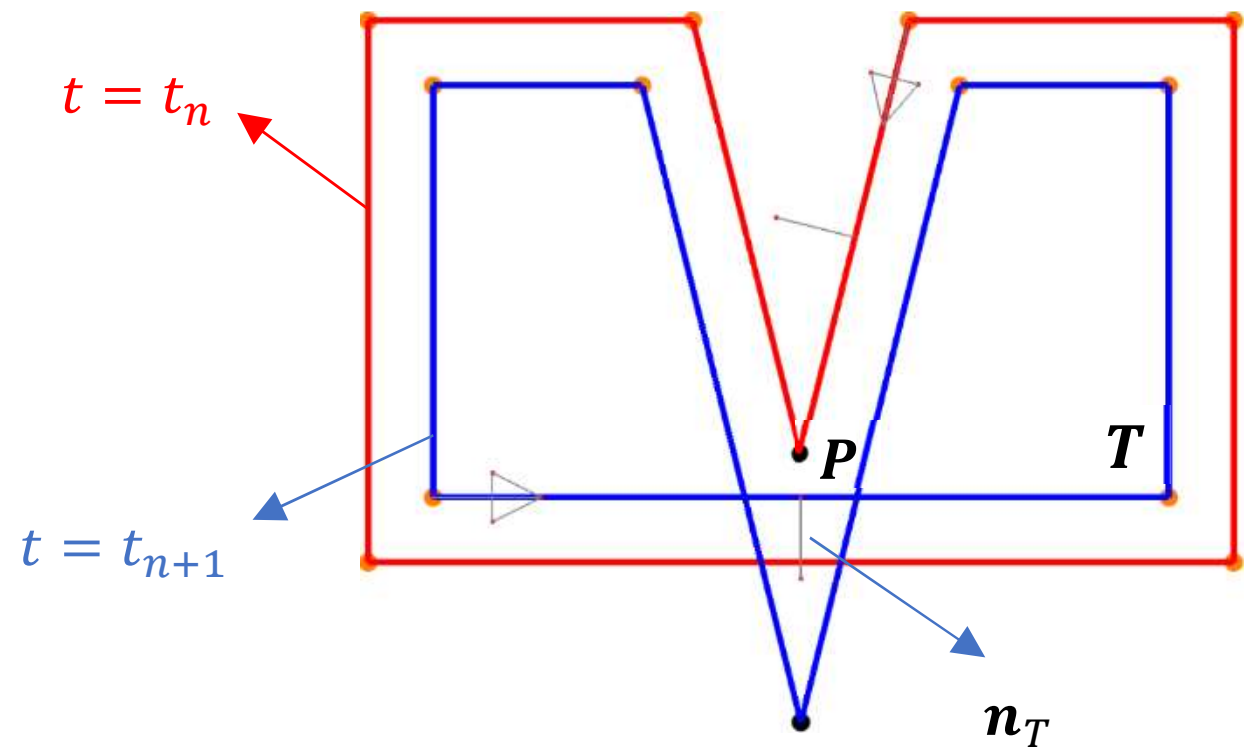

*Fig. 5 Edge penetration*

At the instance of edge split event, vertex $\boldsymbol{P}$ must be on the edge belonging to vertex $\boldsymbol{T}$ (see Fig. 6). Therefore, the following relation must hold:

$$(\boldsymbol{P}_S - \boldsymbol{T}_S) \cdot \boldsymbol{n}_T = 0 \quad (27)$$

Inserting Eq.(26) into Eq.(27), local coordinate corresponding to edge split event is computed as

$$\xi_S = -\frac{(\boldsymbol{P}_M - \boldsymbol{T}_M) \cdot \boldsymbol{n}_T}{(\boldsymbol{P}_D - \boldsymbol{T}_D) \cdot \boldsymbol{n}_T} \quad (28)$$

where $\boldsymbol{P}_M$ and $\boldsymbol{P}_D$ are defined as

$$\boldsymbol{P}_M \coloneqq \frac{1}{2}(\boldsymbol{P}_{n+1} + \boldsymbol{P}_n) \overset{\text{Eq.(17)}}{\triangleq} \boldsymbol{P}_n + \frac{\Delta t_n}{2}\boldsymbol{v}_P$$
$$\boldsymbol{P}_D \coloneqq \frac{1}{2}(\boldsymbol{P}_{n+1} - \boldsymbol{P}_n) \overset{\text{Eq.(17)}}{\triangleq} \frac{\Delta t_n}{2}\boldsymbol{v}_P \quad (29)$$

Definitions for $\boldsymbol{T}_M$ and $\boldsymbol{T}_D$ follows from Eq.(29). Using the equation for the parameterization of the time increment, Eq.(24), time increment corresponding to edge split event is obtained as

$$\Delta t_S = \frac{1}{2}(1+\xi_S)\Delta t_n \quad (30)$$

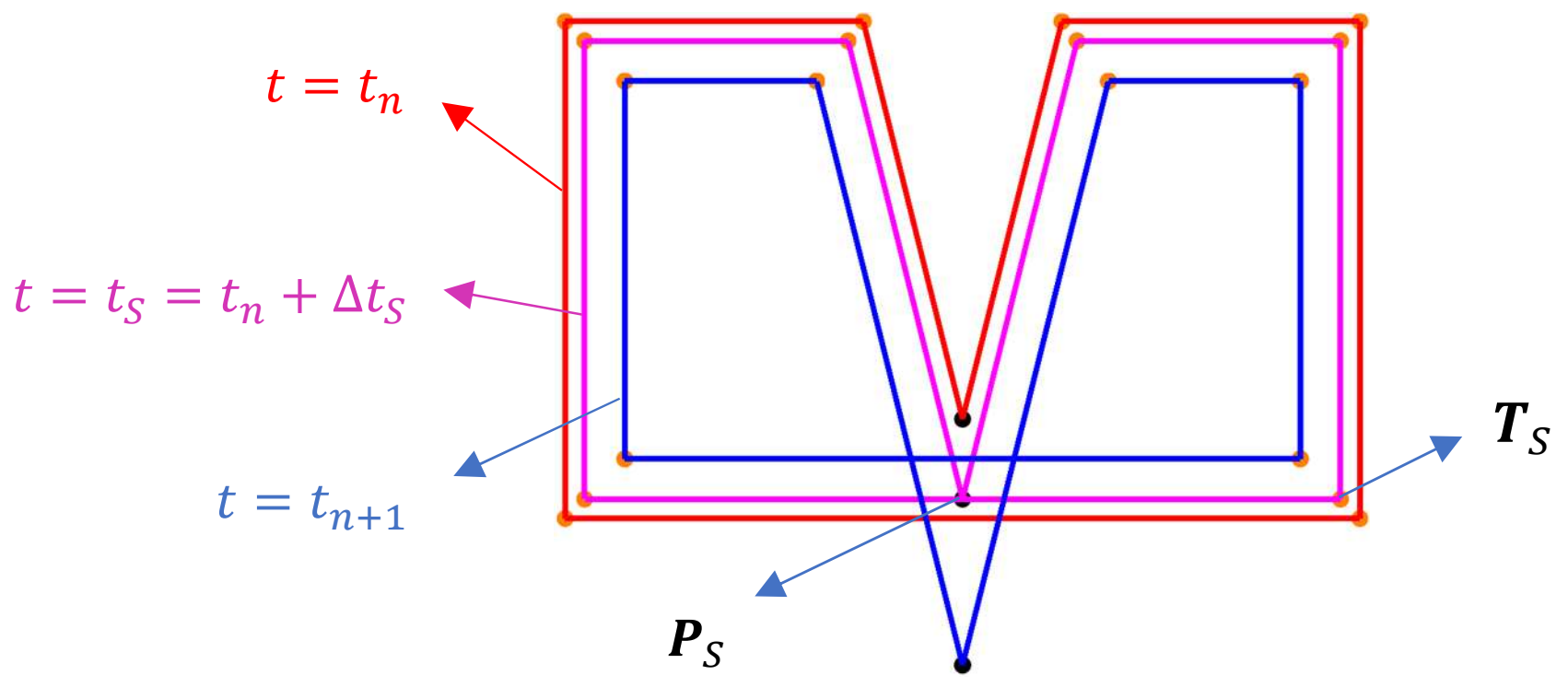

*Fig. 6 The instance of edge split event*

## 3.3 Predictor-corrector algorithm

For a given time increment, predicted position of a typical vertex $\boldsymbol{P}$ is computed by direct integration as

$$\boldsymbol{P}_{n+1}^{pred} = \boldsymbol{P}_n + \Delta t_n \boldsymbol{v}_P \tag{31}$$

After predicted positions are computed, admissibility of the polygon is checked with respect to edge collapse and edge split events using Eq.(21) and Eq.(27). If there is no event taking place, final positions are equal to predicted positions, i.e.,

$$\boldsymbol{P}_{n+1} = \boldsymbol{P}_{n+1}^{pred} \tag{32}$$

If there is an edge collapse event within the time increment, predicted positions are corrected as

$$\boldsymbol{P}_{n+1} = \boldsymbol{P}_{n+1}^{pred} + \Delta \boldsymbol{P}_C \tag{33}$$

with

$$\Delta \boldsymbol{P}_C = (\Delta t_C - \Delta t_n) \boldsymbol{v}_P \tag{34}$$

Similarly, if there is an edge split event within the time increment, predicted positions are corrected as

$$\boldsymbol{P}_{n+1} = \boldsymbol{P}_{n+1}^{pred} + \Delta \boldsymbol{P}_S \tag{35}$$

with

$$\Delta \boldsymbol{P}_S = (\Delta t_S - \Delta t_n) \boldsymbol{v}_P \tag{36}$$

If both edge collapse and edge split events take place within the same increment, the one which happens earlier is considered and the other one is disregarded.

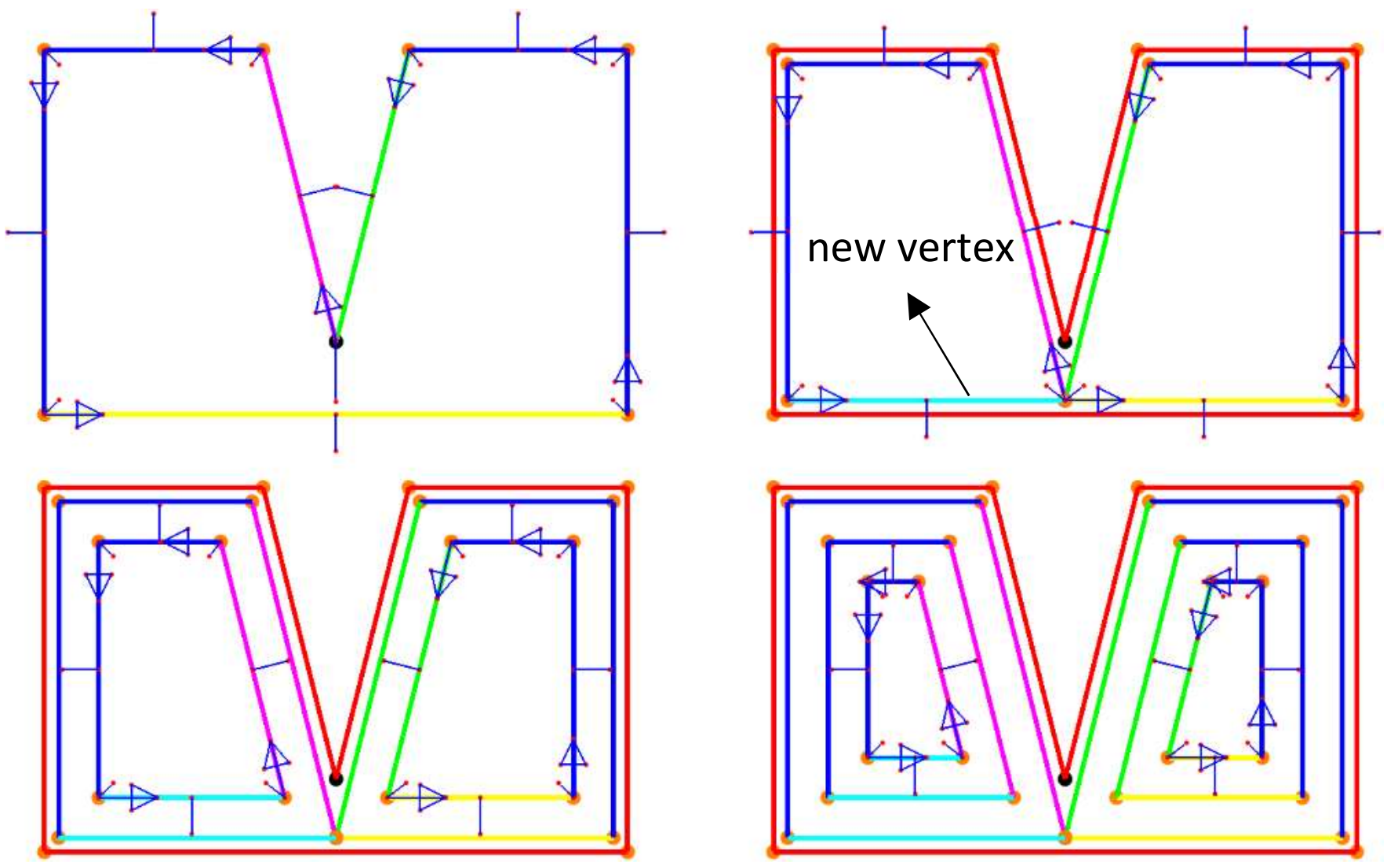

*Fig. 7 Vertex-in-edge collision*

## 3.4 Data structures

Polygons are stored as doubly linked vertex lists. A vertex stores basically information about position, velocity, unit vector to edge before it, normal vector to edge before it, pointer to vertex before it and pointer to vertex after it. To keep track of the evolution of straight skeleton structure, a vertex also stores pointers to the vertex above and the vertex below, which are associated, respectively, with the polygon above and polygon below.

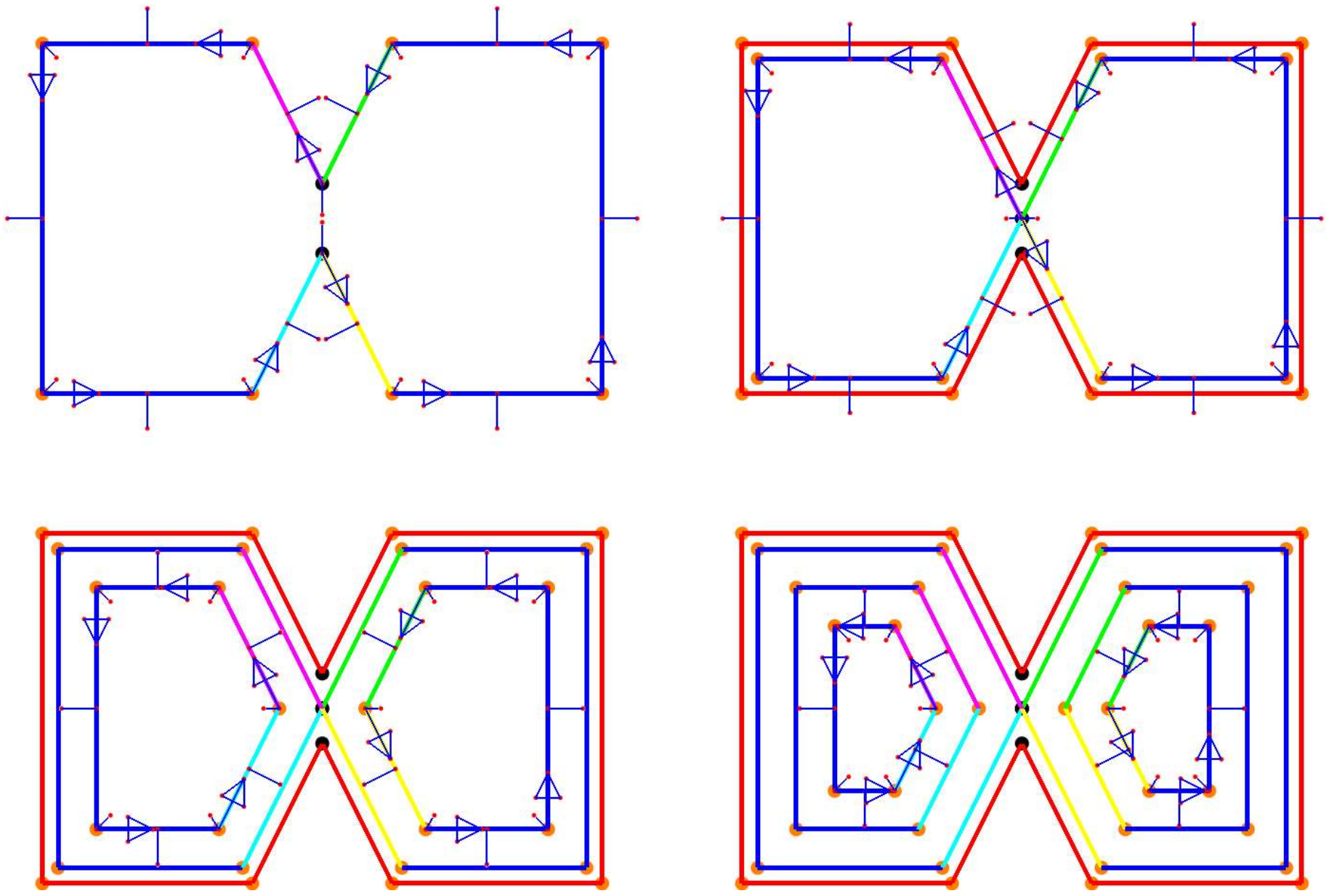

*Fig. 8 Vertex-on-vertex collision*

Data structure described must be adapted as the edge collapse and edge split events occur during incremental motion. When an edge corresponding to a vertex is collapsed, it is simply removed from the linked list and links, namely vertex before and vertex after pointers, for the neighbor vertices are updated. When an edge split event occurs, there are two possibilities. Split event could occur from an edge, vertex-in-edge (see Fig. 7), or from a vertex, vertex-on-vertex (see Fig. 8). Note that in case of vertex-in-edge, a new vertex is created and added to linked list. After the links are updated, velocity of the contributing vertices must be recomputed. In case of simultaneous multiple vertex-on-vertex collision events, links are updated in a recursive manner until no more unprocessed event left (see Fig. 9). Note also that in contrary to [28], no preprocessing is necessary to order the vertices contributing to the simultaneous vertex-on-vertex collisions. Algorithm works perfectly fine when those vertices are handled in an arbitrary order. The ambiguities related with vertex-on-vertex collisions have been discussed in depth in [27, 29]. The link update should work fine for rather extreme cases like simultaneous penetration of an edge multiple times (see Fig. 10).

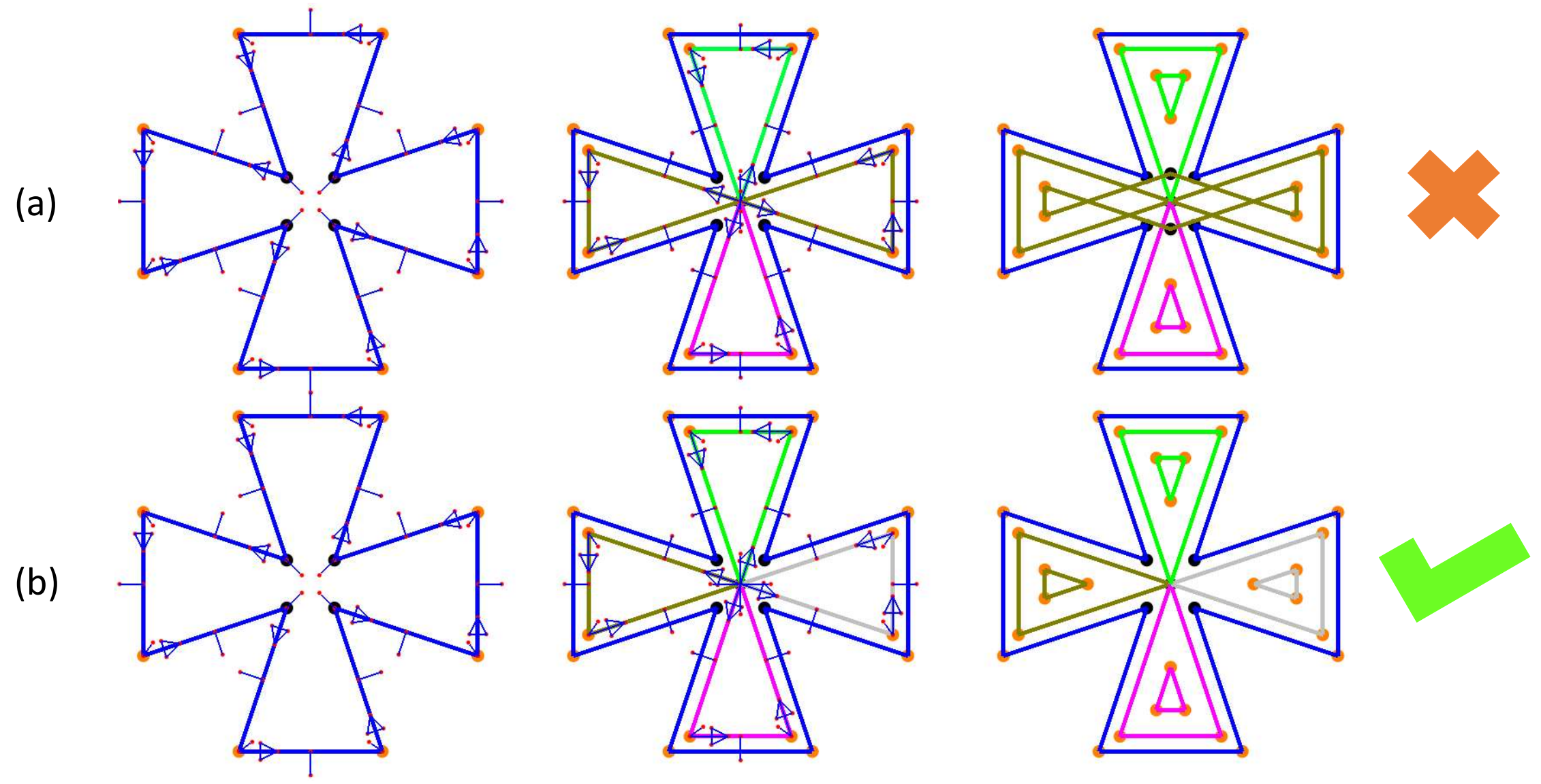

*Fig. 9 Multiple simultaneous vertex-on-vertex edge split event.*
*(a) one-pass resolution. (b) recursive resolution*

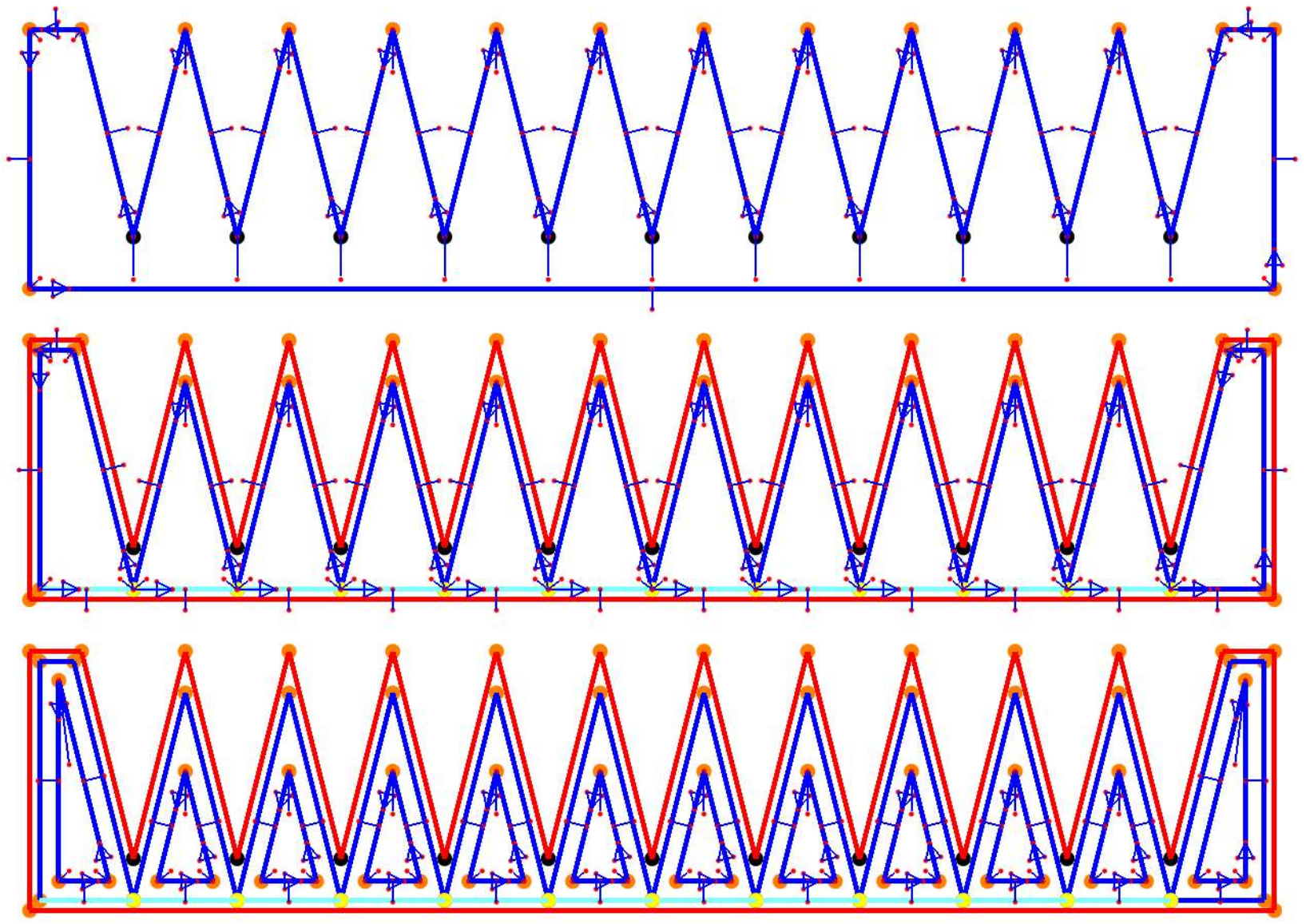

*Fig. 10 Multiple simultaneous vertex-in-edge edge split event*

## 3.5 Search for edge penetration

The most time-consuming part the of the predictor-corrector algorithm is the construction of the vertex pairs for edge penetration within a given time increment. In the search for edge penetration, no differentiation is made between convex and reflex vertices, because depending on the weight both can cause edge split event to occur. A vertex can penetrate into the polygon both from inside and from outside, see Fig. 11. One must be careful because sometimes during a given time increment a vertex could even penetrate through, see Fig. 11. Criteria given by Eq.(27), holds in all possible scenarios and could safely be used. To test Eq.(27) for a given vertex *P*, all the vertices, except *P* and vertex after *P*, must be checked. This brute force approach is not very feasible especially for a polygon with many vertices and independent subpolygons. Instead of performing a brute force search, an enhancement based on bounding box of a vertex or a quadtree structure or any other similar approach, like given in [26], based on spatially grouping the vertex input could be proposed. Speed of the algorithm is not the main concern of the current study and therefore no attempt has been made to accelerate search for edge penetration. Proposed brute force search algorithm works also fine for rather extreme cases where penetrating vertices or penetrated edges are stationary, see Fig. 12.

(a)

(b)

(c)

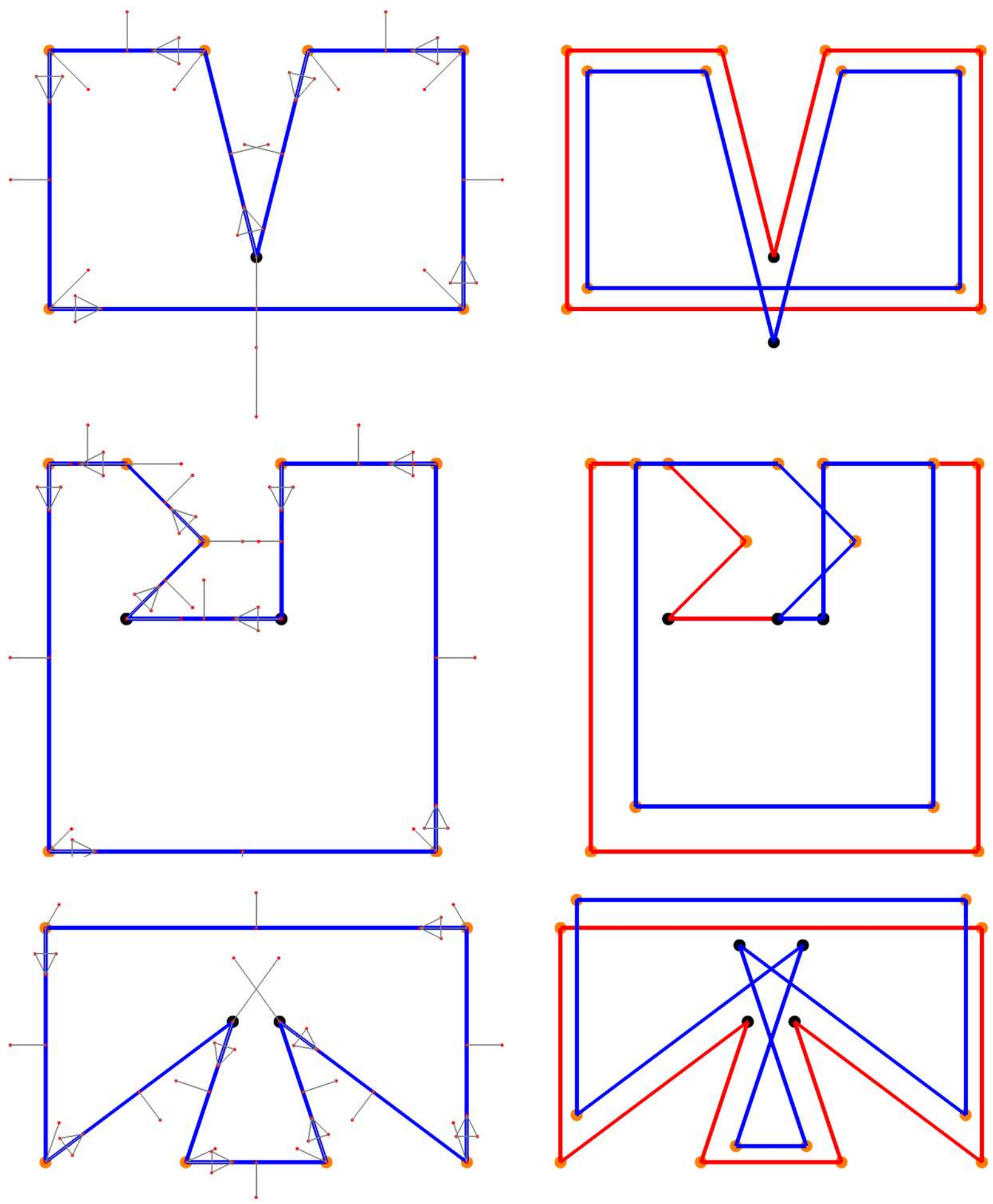

*Fig. 11 Edge penetration. Different possibilities. Vertex penetrates (a) from inside outside, (b) from outside inside, (c) through*

(a)

(b)

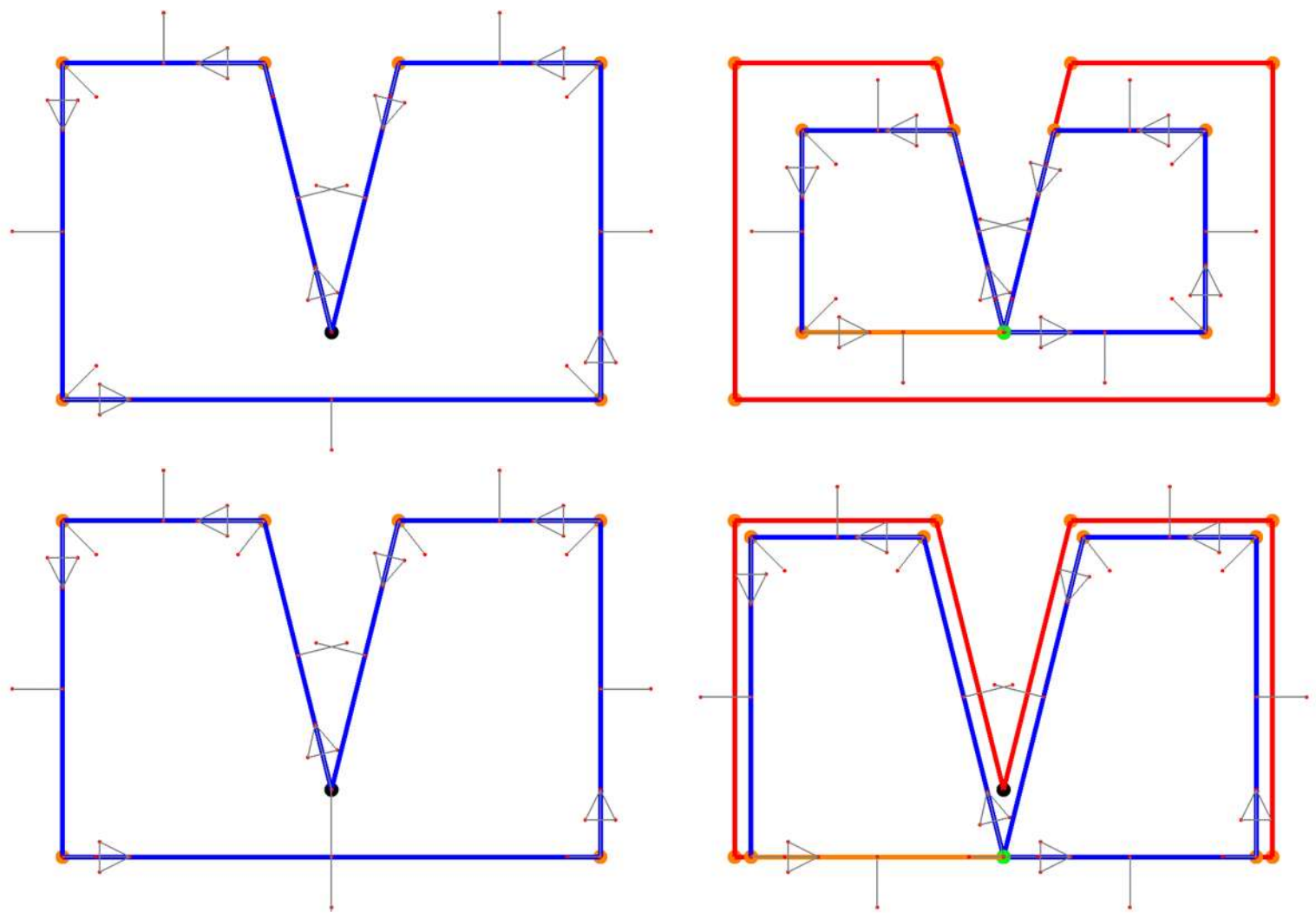

*Fig. 12 Edge penetration search. Extreme cases (a) stationary vertex and moving edge. (b) moving vertex and stationary edge*

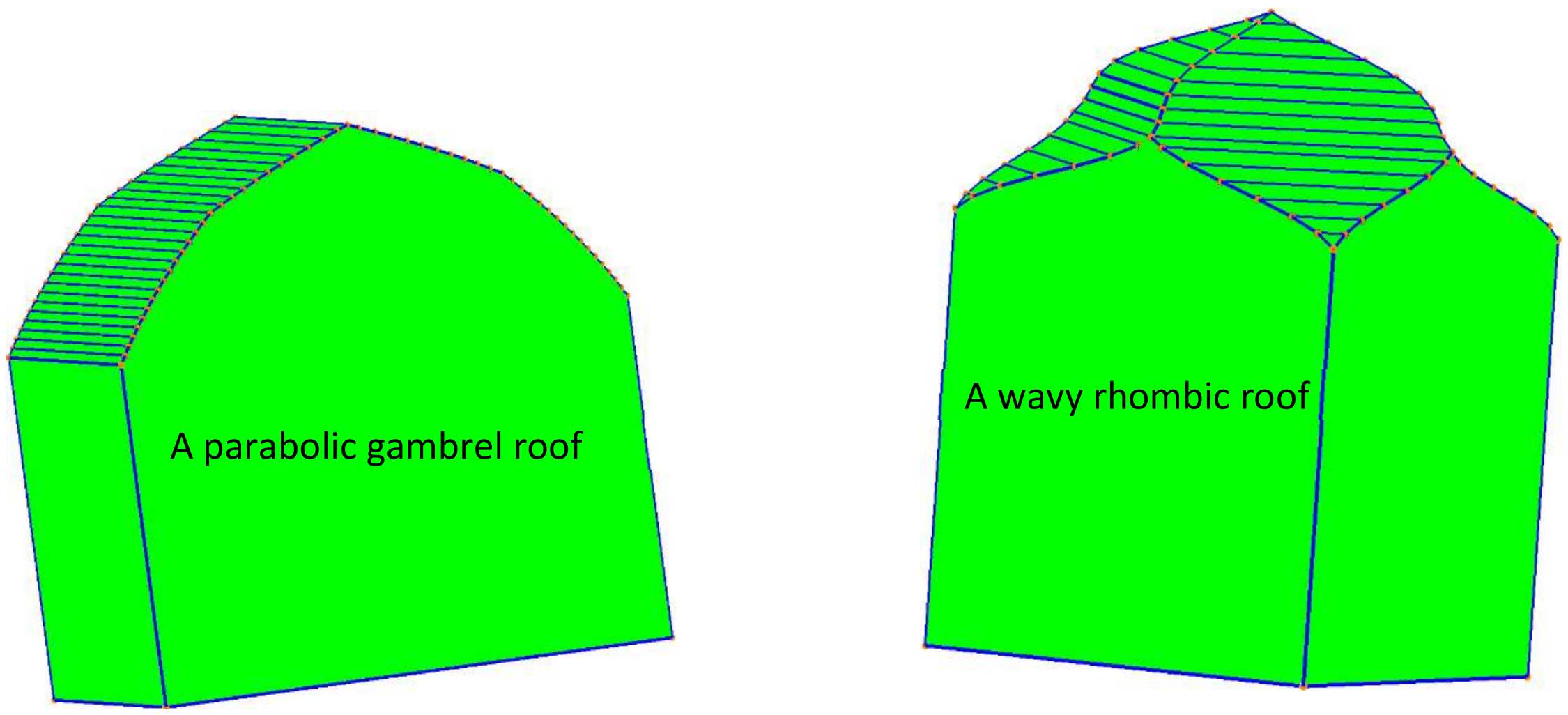

*Fig. 13 Roofs with curved edges*

## 4 Roof modelling

To construct a roof topology using straight skeleton algorithm proposed, a proper expression for the vertex velocity is necessary. For a roof with straight edges, velocity for a typical vertex *P* assumes the form

$$\boldsymbol{v}_P = \overline{\boldsymbol{v}}_P + v_z \boldsymbol{e}_z \quad (37)$$

where $\overline{\boldsymbol{v}}_P$ is the projection of the velocity vector $\boldsymbol{v}_P$ onto the polygon plane which is spanned by unit vectors $\boldsymbol{e}_x$ and $\boldsymbol{e}_y$, and, therefore, has the form

$$\overline{\boldsymbol{v}}_P = v_{Px} \boldsymbol{e}_x + v_{Py} \boldsymbol{e}_y \quad (38)$$

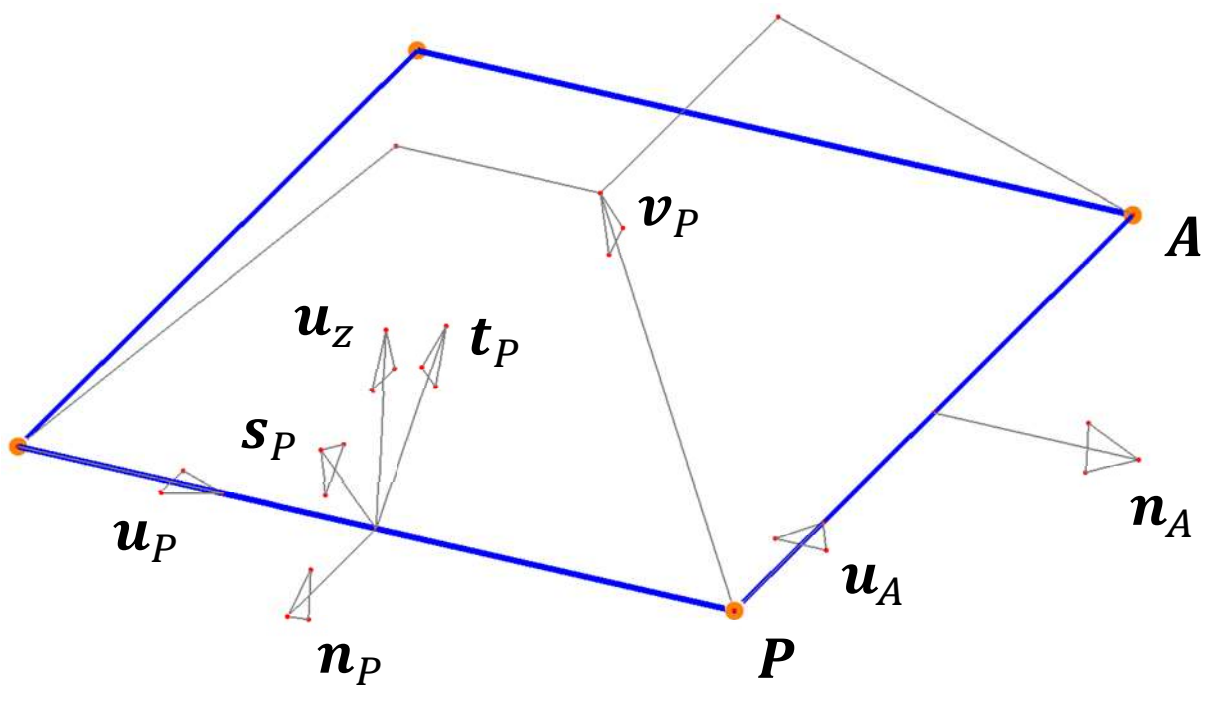

*Fig. 14 Roof planes associated with vertex* P *and* A

Note that *z* component of the velocity vector is same for all the vertices therefore polygon remains planar while deforming. By setting $v_z = 1$, the amount of offset applied to polygon in *z* direction turns out to be equal to the corresponding time increment. One can construct piecewise linear approximations to the roofs with curved edges by connecting $z$ component of the velocity vector to a function (see Fig. 13) without introducing any additional complexity to the underlying algorithm.

For a typical vertex *P*, velocity components $v_{Px}$ and $v_{Py}$ are computed using the inclination angles $\alpha_P$ and $\alpha_A$, with $\alpha \in (0, \pi)$, assigned respectively to the vertex *P* and vertex after it, which we denote as *A* (see Fig. 14 and Fig. 15). The roof plane associated with a typical vertex is oriented with respect to surface normal defined as $\boldsymbol{s}_P$. To compute $\boldsymbol{s}_P$ consistently, we first define a unit vector $\boldsymbol{t}_P$ which lies inside the roof plane as

$$\boldsymbol{t}_P = \cos(\pi - \alpha_P)\,\boldsymbol{n}_P + \sin(\pi - \alpha_P)\boldsymbol{u}_z \tag{39}$$

where $\boldsymbol{u}_z$ is unit vector normal to the polygon plane. Both $\boldsymbol{t}_P$ and $\boldsymbol{u}_P$ lie inside roof plane associated with the vertex *P*. Then normal to this roof plane can easily be constructed by taking the cross product as

$$\boldsymbol{s}_P = \boldsymbol{u}_P \times \boldsymbol{t}_P \tag{40}$$

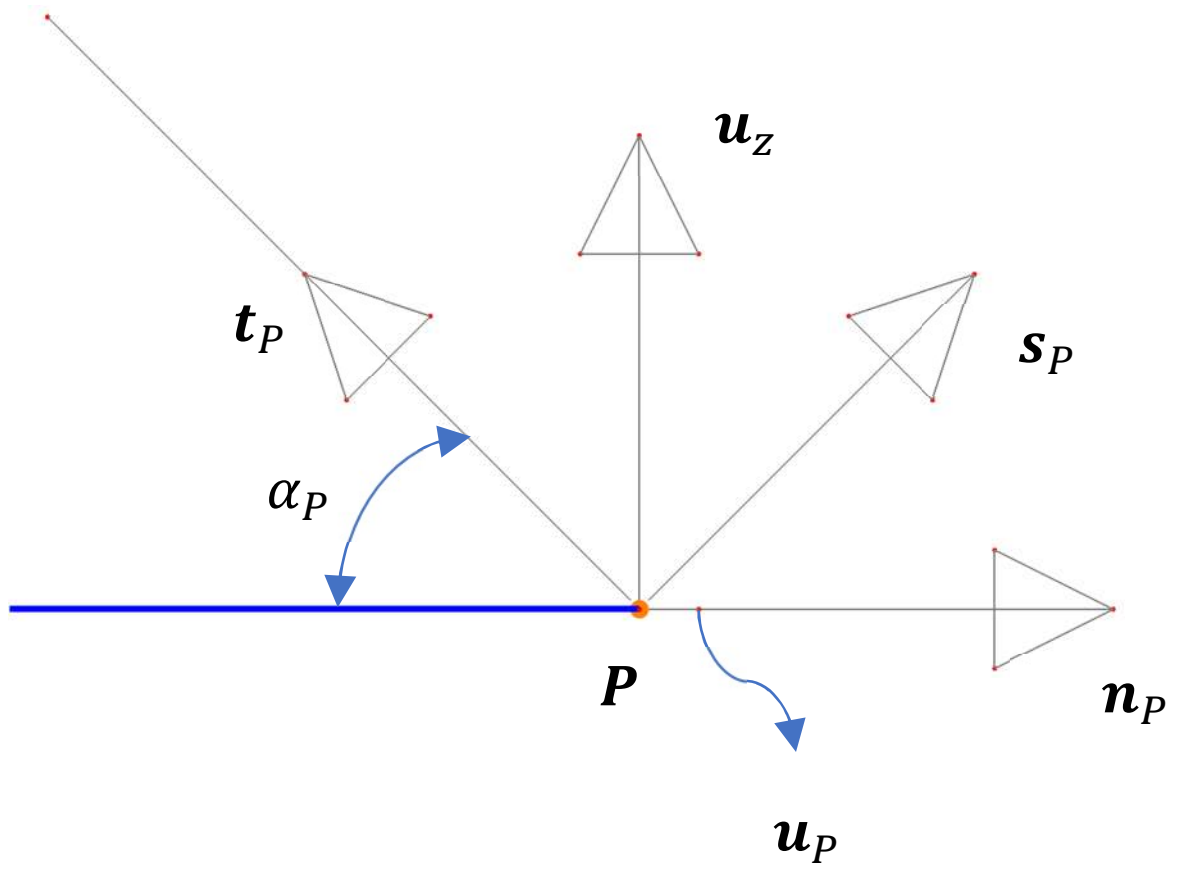

*Fig. 15 Construction of roof unit normal $\boldsymbol{s}_P$*

While the polygon is deforming vertices must lie inside the roof surfaces belonging to them, which can be expressed by the following set of linear equations:

$$\begin{aligned} \boldsymbol{s}_P \cdot (\boldsymbol{P}_{n+1} - \boldsymbol{P}_n) &= 0 \\ \boldsymbol{s}_A \cdot (\boldsymbol{P}_{n+1} - \boldsymbol{P}_n) &= 0 \end{aligned} \tag{41}$$

Note that computation of $\boldsymbol{s}_A$ follows from Eq.(39) and Eq.(40). With the insertion of Eq.(17) and Eq.(38), Eq.(41) can be rewritten as

$$\begin{aligned} \bar{\boldsymbol{s}}_P \cdot \bar{\boldsymbol{v}}_P + s_{Pz} \cdot v_z &= 0 \\ \bar{\boldsymbol{s}}_A \cdot \bar{\boldsymbol{v}}_P + s_{Az} \cdot v_z &= 0 \end{aligned} \tag{42}$$

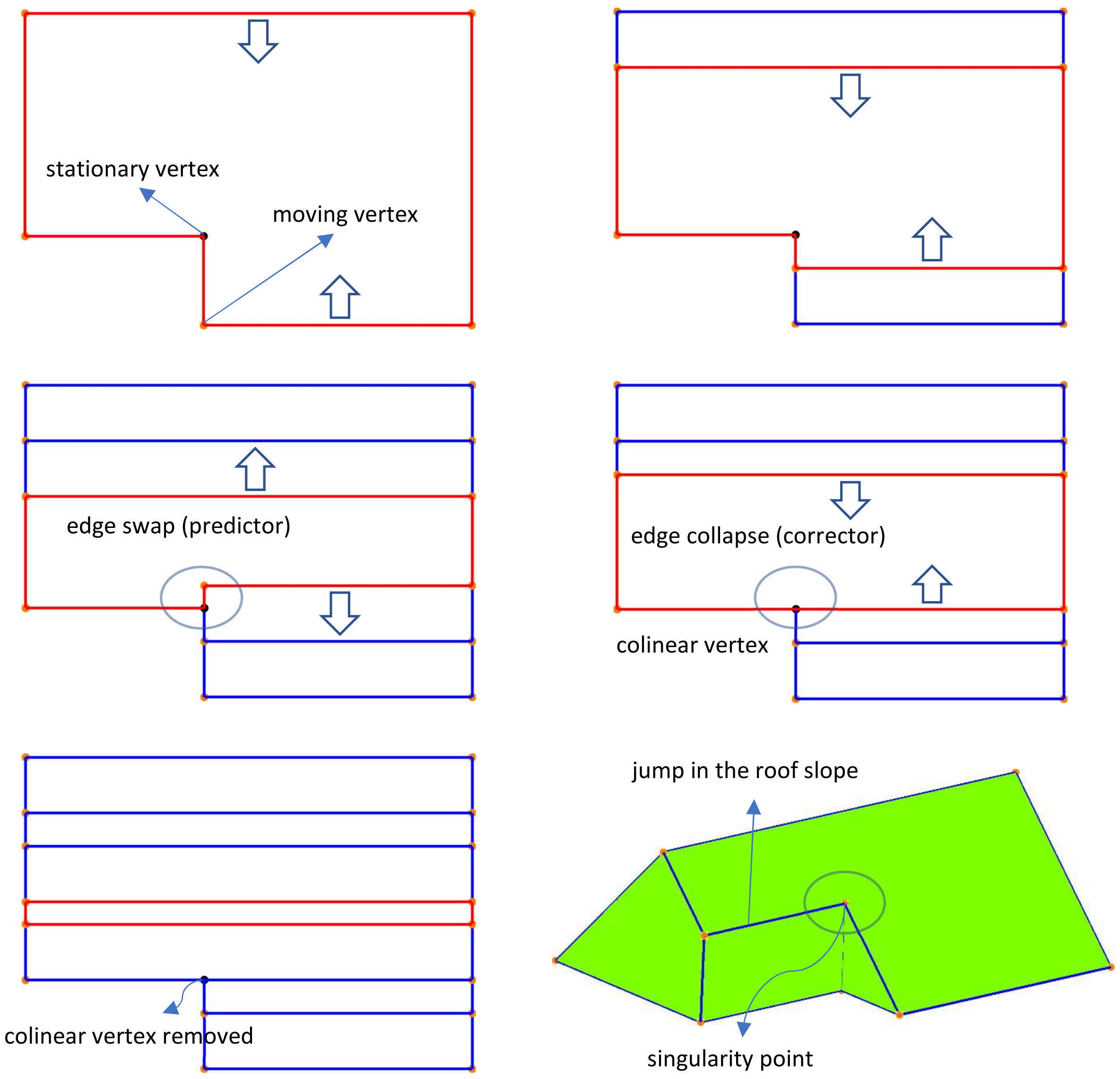

*Fig. 16 Singularity due to colinear vertex*

Expressions with bar follows the definition given in Eq.(38). If the projections $\bar{\mathbf{s}}_P$ and $\bar{\mathbf{s}}_A$ are not colinear, Eq.(42) defines two linearly independent equations for unknown velocity components $v_{Px}$ and $v_{Py}$ and hence can easily be solved. Note that if, for example, stationary vertices are present, one can easily end up with an edge collapse event where colinear vertices emerge during incremental motion, see Fig. 16. For colinear vertices, the set of equations given by Eq.(42) becomes automatically linearly dependent and as a result solution is undefined. Therefore, after an edge collapse event, colinear vertices must be found and removed from the polygon. In the proposed algorithm, in the auto mode, the part belonging to removed vertex is added to the edge belonging to the vertex after it. However, due to underlying incremental nature, one has the capability to interrupt and make manual adjustment to decide how the base polygon will propagate further. Alternative treatments of the colinearity issue are discussed in detail in [27, 28]. Independent of the solution proposed, there will be a jump in the velocity vector along a certain part of the polygon after the colinear vertices are removed. In roof modelling, it can be interpreted as sudden change in the slope, see Fig. 16, which is perfectly sound from geometrical point of view. Note that due to its incremental nature, one could interact with the algorithm at any time and , for example, change slope along some edges or insert and remove vertices in order to construct roofs with custom geometry as given in Fig. 18.

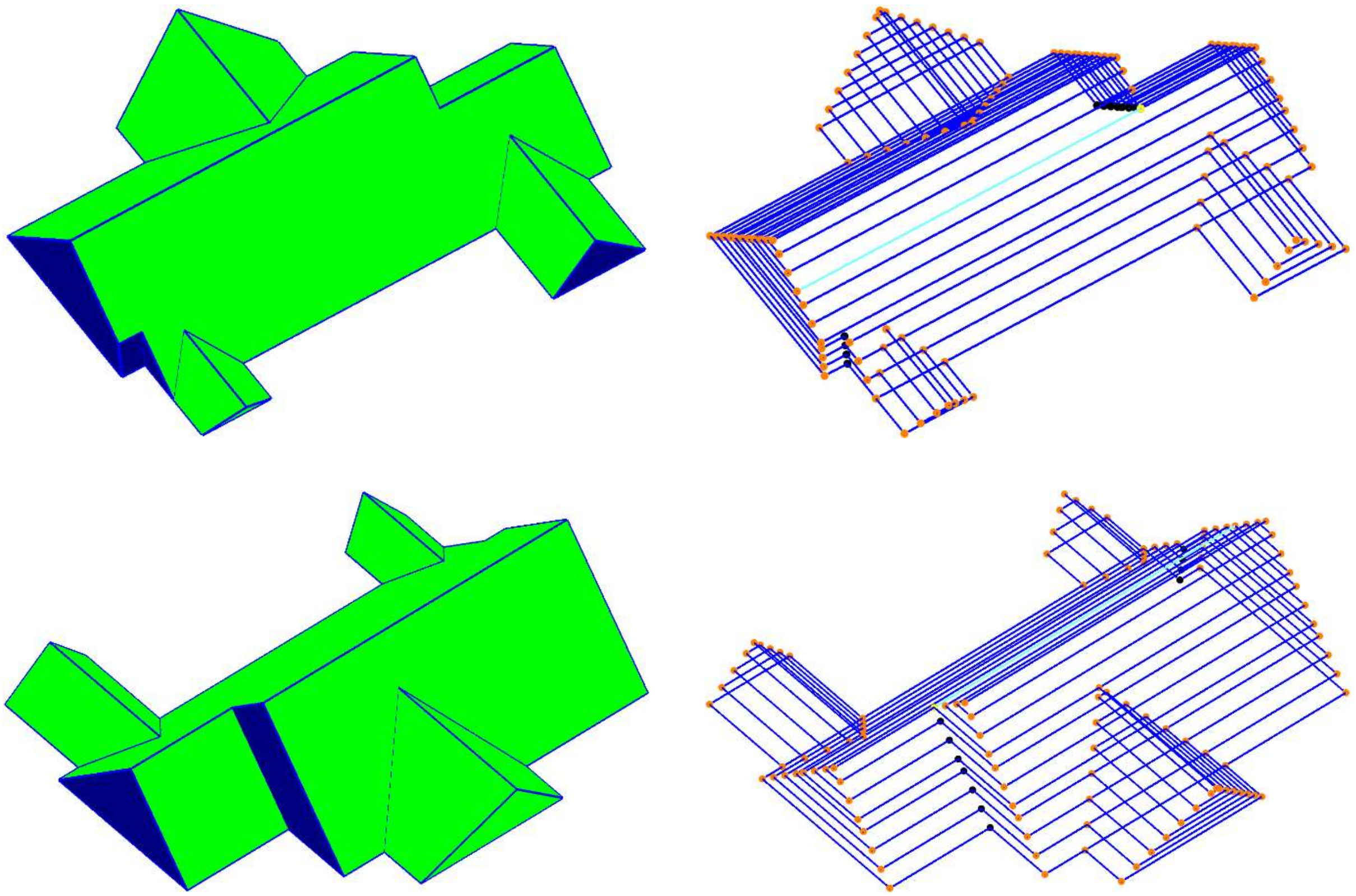

*Fig. 17 A composite roof made up of a polygon with four independent subpolygons*

In the roof construction using weighted straight skeleton algorithm, new position of the polygon is computed based on predictor-corrector algorithm for a prescribed offset in *z* direction. A roof is complete if there remains no possibility to further apply an offset. Recall that a polygon cannot be offset if it has collapsed down to a single vertex or a single edge, two vertices.  Offset polygons are kept and used to construct the roof polygons, namely the straight skeleton structure, in a systematic way by paying special attention to the vertices which are split, collapsed or colinear. In Fig. 17 and Fig. 18, several representative non-exhaustive examples are given.

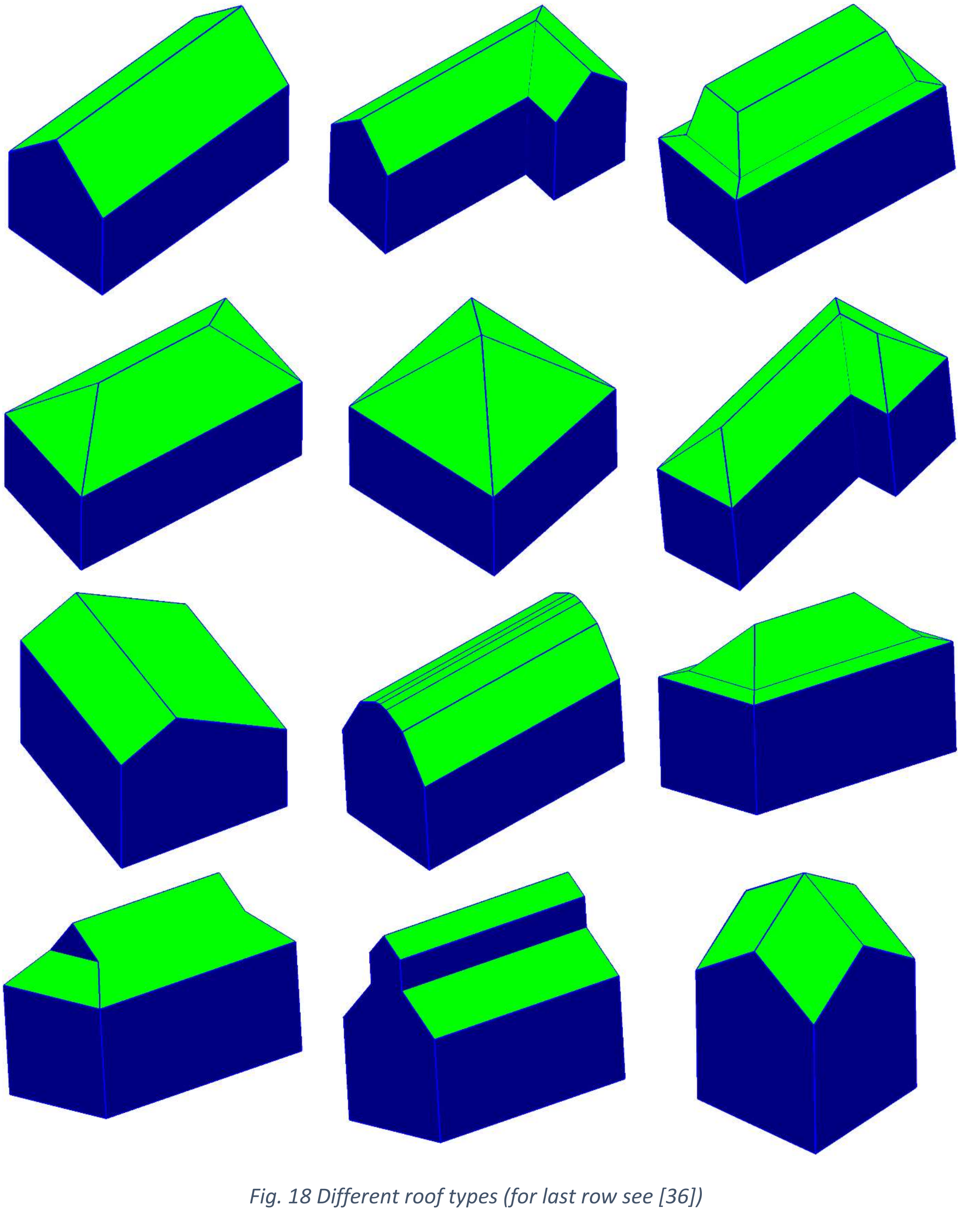

*Fig. 18 Different roof types (for last row see [36])*

# 5 Straight Skeleton Development Kit

In order to develop a reliable general purpose straight skeleton algorithm, or algorithm of any kind, it is essential to have a tool equipped with visualization capabilities. Therefore, a graphical user interface has been developed by the author in parallel for the construction of new predictor-corrector type incremental weighted straight skeleton algorithm which is based on the notion of deforming polygon using C++ programming language, see Fig. 19. With the help of this tool, it was possible to test each individual building block of the algorithm separately, which gave tremendous amount of speed up to converge to the current version of the algorithm. Program was not only used to develop the algorithm but also to automize repetitive error prone and time-consuming tasks like preparing the figures for the work presented here.

# 6 Conclusion

A universal predictor-corrector type incremental algorithm for the construction of weighted straight skeletons of 2D general planar polygons, possibly with holes, has been developed based on the notion of deforming polygon. An algorithm of this class has never been used before for straight skeletons. Algorithm resembles the techniques which are commonly used in computational solid mechanics. The algorithm works with the raw input provided by the underlying polygon. Additional tools like kinetic triangulation or motorcycle graphs are not needed. Polygon edges which move at different speeds and/or start to move at different times are supported by default. In the algorithm predicted positions of the vertices are computed by direct time integration first. Then admissibility of the polygon is checked with respect to edge swap and edge penetration. Edge swap and edge penetration are used to detect, respectively, edge collapse and edge split events within an increment. For a given event, corrected position of the vertices is computed exactly by using linear interpolation. In the algorithm, there is no special treatment for the reflex vertices. Positively, or negatively, weighted and stationary vertices are dealt within the same framework proposed. By assuming a special form for the vertex velocity, the algorithm has been used to construct roof topology starting from a floor plan given. The algorithm could also be used to generate piecewise approximations to the roofs with curved edges by simply connecting $z$ component of the vertex velocity to a prescribed function without introducing any additional complexity. A graphical user interface, Straight Skeleton Development Kit, has been developed in parallel from scratch to construct, improve and test the algorithm.

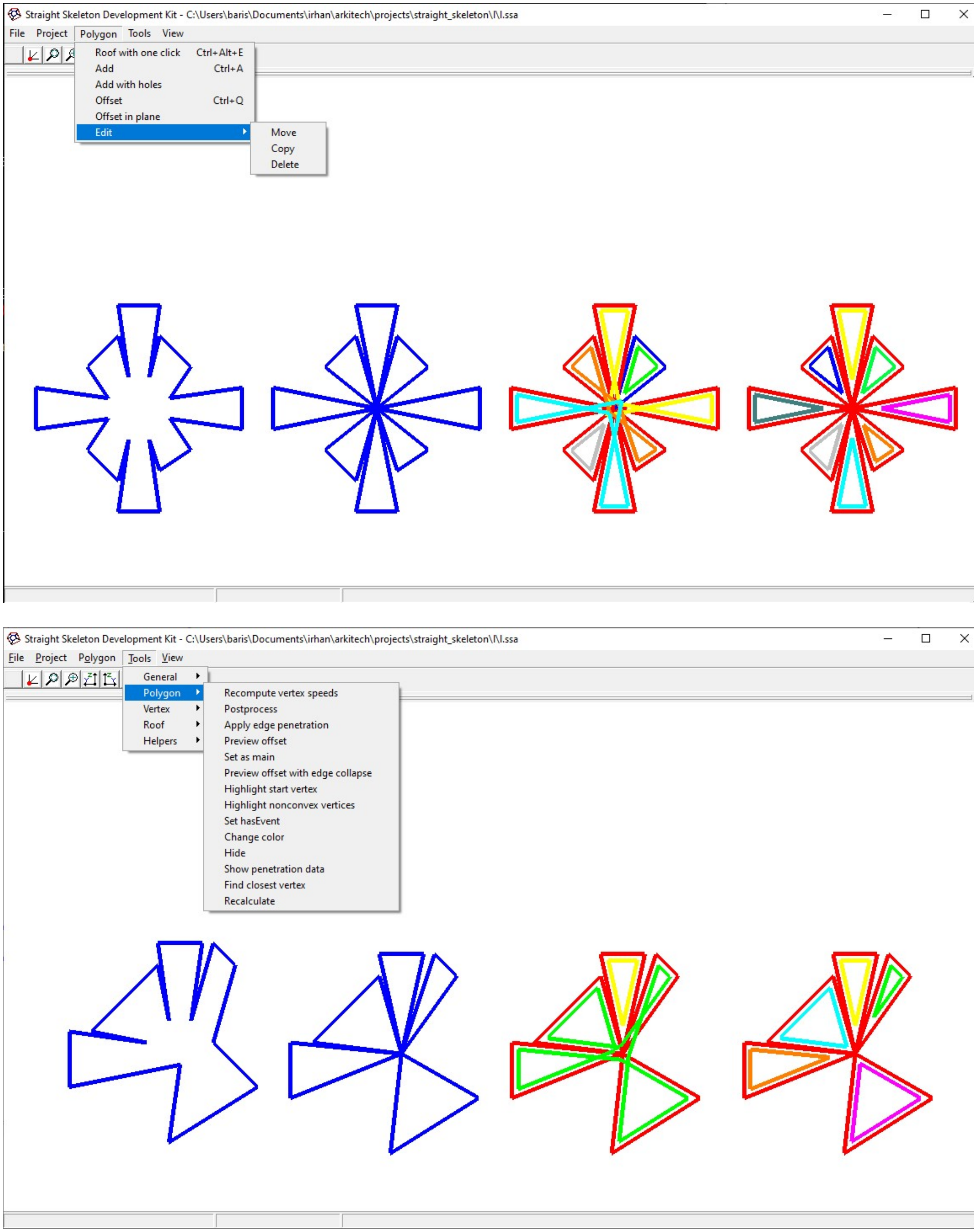

*Fig. 19 Straight Skeleton Development Kit*